\documentclass[submission,copyright,creativecommons]{eptcs}
\providecommand{\event}{QAPL 2016} 
\usepackage{breakurl}             

\newcommand{\vspaceabovesection}{}
\newcommand{\vspaceaboveformulas}{}
\newcommand{\vspacebelowformulas}{}
\newcommand{\vspacebelowtables}{}

\usepackage{cite}
\usepackage{framed}
\usepackage{tabu}
\usepackage[framemethod=tikz]{mdframed}
\usepackage{float}
\usepackage{calc}
\usepackage{pgfplots}
\usepackage{tikz}
\usepackage{sidecap}
\usepackage{fancyvrb}
\usepackage{epsfig}
\usepackage{wrapfig}
\usepackage{subfig}
\usepackage{verbatim}
\usepackage{amsfonts}
\usepackage{enumitem}
\usepackage{amsmath}
\usepackage{color}
\usepackage{graphicx}
\usepackage{amssymb}
\usepackage{listings}

\title{Evaluating load balancing policies\\ for performance and energy-efficiency}
\author{
	Freek van den Berg \thanks{ This research was supported as part of the Dutch national program COMMIT, and carried out as part
of the Allegio project under the responsibility of the Embedded Systems Innovation group of TNO, with
Philips Medical Systems B.V. as the carrying industrial partner.}
	\institute{University of Twente\\
		Enschede, The Netherlands}
	\email{\quad f.g.b.vandenberg@utwente.nl}
	\and
	 Bj\"{o}rn F. Postema \thanks{The work in this paper has been supported by the Dutch national STW project Cooperative Networked Systems (CNS), as part of the program ``Robust Design of Cyber- Physical Systems'' (CPS).}
	\institute{University of Twente\\
		Enschede, The Netherlands}
	\email{\quad b.f.postema@utwente.nl}
	\and
	 Boudewijn R. Haverkort \footnotemark[1] \ \footnotemark[2]
	\institute{University of Twente\\
		Enschede, The Netherlands}
	\email{\quad b.r.h.m.haverkort@utwente.nl}	
}
\def\titlerunning{Energy-efficient load balancing policies}
\def\authorrunning{F. van den Berg, B.F. Postema \& B.R. Haverkort}
\def\copyrightholders{van den Berg, Postema \& Haverkort}
\begin{document}
	\maketitle

\def\lastname{Van den Berg}
\begin{abstract}
Nowadays, more and more increasingly hard computations are performed in challenging fields like weather forecasting, oil and gas exploration, and cryptanalysis. Many of such computations can be implemented using a computer cluster with a large number of servers. Incoming computation requests are then, via a so-called load balancing policy, distributed over the servers to ensure optimal performance. Additionally, being able to switch-off some servers during low period of workload, gives potential to reduced energy consumption. Therefore, load balancing forms, albeit indirectly, a trade-off between performance and energy consumption.
In this paper, we introduce a syntax for load-balancing policies to dynamically select a server for each request based on relevant criteria, including the number of jobs queued in servers, power states of servers, and transition delays between power states of servers.
To evaluate many policies, we implement two load balancers in: (i) iDSL, a language and tool-chain for evaluating service-oriented systems, and (ii) a simulation framework in AnyLogic. Both implementations are successfully validated by comparison of the results.
\end{abstract}


\newcommand{\xx}{\ | \ }
\newcommand{\newpagebeforechapter}{}


\vspaceabovesection
\section{Introduction}
\label{sec:introduction}
In 2006, Al Gore created a global awareness and willingness to reduce greenhouse gasses by releasing his film ``An inconvenient truth'' \cite{nolan2010inconvenient}. Roughly one third of these green house gasses are caused due to the generation of electricity\footnote{http://www3.epa.gov/climatechange/ghgemissions/sources.html}. Additionally, approximately 1.1\%-1.5\%  of the worldwide electricity was consumed by data centres in 2010 \cite{Koomey2011}. In 2012, Neelie Kroes \cite{Kroes2012}, former vice-president of the European Commission responsible for the digital agenda, states that ICT consumes 8\% to 10\% of all European electricity, which is approximately the total power consumption of whole of the Netherlands. All of the above has led to a stronger focus on green ICT solutions in which saving electricity has becoming increasingly important.

A major energy consumer in ICT are server farms. These server farms consist of many servers that take care of some load. A server environment is set up in a data centre, which provides the infrastructure that enables servers to function. These data centres often consist of many more components that consume energy. Because the energy consumption of these components positively correlates with power consumed by the servers, even small improvements made at server level have a strengthened effect (the so-called ``cascade-effect'' \cite{ENPower2009}). 

One of the ways to reduce energy consumption is achieved with the aid of power management features. \textit{Power management} allows to switch between power states of servers to reduce power consumption, while trying to keep the performance intact (e.g., bringing to sleep underutilised servers). There are two key elements that construct a power management \textit{policy}, namely: (i) \textit{strategies} and (ii) \textit{load balancing}. First, power management strategies describe when servers should switch between the power states. Second, the load should be balanced among the servers such that optimal performance is obtained.

This paper proposes a powerful, yet concise, policy language, which covers, among others, strategies that observe the size of the queue to decide to which server jobs are assigned. Furthermore, servers are put to sleep when idling with a simple time-out mechanism, which should be easy to implement in actual servers as literature suggests \cite{845896}. The method proposed in this paper allows us to explore a large set of designs by adjusting only three parameters: (i) \textit{queue size threshold}, (ii) \textit{idle time-out}, and (iii) \textit{non-determinism resolution}. In the end, this leads to interesting power-performance trade-offs, i.e., we explore the possibility to exchange reduction of power consumption at the cost of performance.

The policies are implemented as extensions to iDSL and AnyLogic so that policies can be automatically evaluated. This provides insights in  the effectiveness of the policies with respect to performance and energy consumption. Also, both implementations are validated by comparison of the results.



\noindent In this paper, we answer the following main research question:\\
\ \\
\centerline{How to obtain high-performance, energy-efficient load balancing policies?}\\
\ \\ 
\noindent The combined answers to the following research questions answer this main research question.
\begin{enumerate}[label=$\mathcal{K}$\arabic{enumi}]
\item  What constitutes a formal model of a load balancer? \label{sec:research_question_formal_model}
\begin{enumerate}[label=$\mathcal{K}$\arabic{enumi}\alph{enumii}]
\item How to model the performance and energy of a load balancer? \label{sec:research_question_formal_model_perf_energy}
\item How to model a load balancer policy? 
\label{sec:research_question_formal_model_lb_policy}
\end{enumerate}
\item How to automatically compare the performance and energy of many designs? 
\begin{enumerate}[label=$\mathcal{K}$\arabic{enumi}\alph{enumii}]
\item How to evaluate many designs to find good policies regarding performance and energy consumption?\label{sec:research_question_idsl}
\item How to validate the evaluated results of the performance and energy?
\label{sec:research_question_anylogic}
\end{enumerate}
\end{enumerate}

\vspace{2mm}\noindent This paper is further organised as follows. 
Section \ref{sec:system_description} provides the system description of a load balancer.
Section \ref{sec:formal_model} formalises a load balancer (question \ref{sec:research_question_formal_model}), including an energy and performance model (question \ref{sec:research_question_formal_model_perf_energy}), as well as a load balancer policy model (question \ref{sec:research_question_formal_model_lb_policy}).
Section \ref{sec:two_implementations} then provides two implementations of a load balancer, using iDSL and Anylogic respectively, which are both used to evaluate many different designs (question \ref{sec:research_question_idsl}).
The results of these evaluations are then compared in Section \ref{sec:expresults} to assess their validity (question \ref{sec:research_question_anylogic}).
Finally, Section \ref{sec:conclusions_and_future_work} concludes the paper.

\newpagebeforechapter
\vspaceabovesection
\section{System description of load balancers}
\label{sec:system_description}
In this section, a system description of load balancers is provided. Section~\ref{sec:stakeholders} introduces the stakeholders and their priorities. Section~\ref{sec:case_study_general_outline} presents a data centre performance cluster, which is then simplified by introducing a scope and assumptions. Section~\ref{sec:effectiveness_load_balancing_policy} provides properties that constitute an effective policy.

\vspaceabovesection
\subsection{Stakeholders and their concerns}
\label{sec:stakeholders}
In general, load balancing distributes workload among various computing resources, e.g., computers, network links or processing units. Especially server clusters designed for computing have dedicated equipment for balancing load among its servers. This equipment is then often allocated in a data centre, which is a facility used to house computer systems and associated components. 

From the perspective of the data centre infrastructure suppliers we distinguish two major qualities: (i) \textit{customer demands} and (ii) \textit{supplier demands}. According to \cite{Arregoces2003}, the system architecture in data centres is mostly driven by five customer demands: availability, scalability, flexibility, security and performance. Insight into these demands is essential for the quality of the data centre. Infrastructure suppliers, however, focus mainly on \textit{energy consumption}, \textit{IT equipment value} and \textit{staff required}. The total energy consumed in a data centre depends on the energy consumed by its switching gear, batteries, power distribution, servers, chillers, coolers, network equipment and monitoring and control devices.

In computing, performance is the most essential demand for customers and energy consumption of these often very expensive server clusters is really high. So especially in the case of computing, a smart load balancer in combination with power management features offer great opportunities to reduce energy consumption, while performance is kept intact. 

\vspaceabovesection
\subsection{Data centre performance cluster}
\label{sec:case_study_general_outline}
The work in this paper is inspired by the so-called Peregrine cluster at the Center for Information Technology \footnote{Center for Information Technology, http://www.rug.nl/society-business/centre-for-information-technology/} (CIT) in Groningen, the Netherlands, as follows. Assume the CIT decides to actively use power management features for their Peregrine cluster \footnote{Peregrine HPC cluster, https://redmine.hpc.rug.nl/redmine/projects/peregrine} in combination with load balancing. This cluster has a total of 4368 cores with three types of nodes, namely: (i) 162 standard nodes with 2 $\times$ 24 Intel Xeon 2.5 GHz cores; (ii) 6 standard nodes equipped with accelerator cards; and (iii) 7 fast nodes with 4 $\times$ 48 Intel Xeon 2.6 GHz cores. Each standard node consumes approximately 40 W for only the CPU cores\footnote{Intel Ark, http://ark.intel.com/}.


The system of this case study is too complex and knowledge is missing to make a valid statement about performance and energy characteristics. For these reasons, we model the system using the following assumptions. ($\mathcal{A}1$ - $\mathcal{A}10$):


\begin{enumerate}[label=$\mathcal{A}$\arabic{enumi}]
\item Incoming requests arrive according to a Poisson process with a negative exponential distribution with rate 1 request per second.
\item The system consist of four similar resources (or server nodes).
\item A load balancing policy specifies how incoming requests are distributed over these four servers.
\item Servers have four power states each; they are either switched on (``stateOn"), asleep (``stateSleep"), moving from ``stateOn" to ``stateSleep" (``stateSuspend"), or moving from ``stateSleep" to ``stateOn" (``stateWake").
\item A server can only process tasks in state ``stateOn".
\item Servers spend exactly 10 seconds in transition states ``stateSuspend" and ``stateWakeup" each, when changing states.
\item Servers each consume 200 Watt in states ``stateOn", ``stateSuspend" and ``stateWakeup", and 14 Watt in state ``stateSleep", based on empirical studies \cite{Barroso2007,Gandhi2012}
\item Servers have infinite queues that adhere to a non-preemptive FIFO scheduling policy.
\item Servers process incoming requests deterministically with rate 1 request per second.
\item Evaluating a load balancing policy takes no time.
\end{enumerate}

\vspaceabovesection
\subsection{Effectiveness of a load balancing policy}
\label{sec:effectiveness_load_balancing_policy}
The effectiveness of a load balancing policy can be seen as a trade-off between performance properties and power properties. 

\subsubsection{Performance} A policy distributes incoming service requests over a number of servers. The way of distributing strongly affects multiple performance metrics, e.g., the queue sizes and utilization of a specific resource are generally high when the load balancer distributes many requests to the same server. In turn, this increases the latencies for requests that are processed by this server.
In this paper, only the (average) latency is considered, because it is an interesting and easy to understand metric for the customer.

\vspaceabovesection
\subsubsection{Energy consumption} The way a policy distributes incoming service requests indirectly affects energy consumption, viz., when a policy does not distribute requests to any server for a specified amount of time, the server will go to sleep and use only a fraction of energy. In this paper, we consider the average amount of energy per second (in Watt) the four servers uses {\em together}.

\newpagebeforechapter
\vspaceabovesection
\section{A performance and energy model for load balancers}
\label{sec:formal_model}

In the previous section we described load balancers. Here we provide a model to evaluate the performance and energy consumption of load balancing policies. Section~\ref{sec:formal_model_performance_energy} defines performance and energy consumption by considering incoming requests, power states and transitions, and latencies of requests.
Section~\ref{sec:formal_model_policy} specifies load balancer policies using a grammar and semantics, some typical examples, mechanisms to resolve non-determinism, and the design space.

\vspaceabovesection
\subsection{Specifying the performance and energy consumption of a load balancer}
\label{sec:formal_model_performance_energy}
We define the performance and energy consumption of load balancers in three steps. 
Section~\ref{sec:formal_model_performance_energy_incoming_requests}
considers incoming requests and their distribution over resources. Section~\ref{sec:formal_model_performance_energy_power_states} shows how power states and transitions depend on incoming requests.
Section~\ref{sec:formal_model_performance_energy_power_latencies} specifies latencies of requests.

\vspaceabovesection
\subsubsection{Incoming requests and their distribution over the resources}
\label{sec:formal_model_performance_energy_incoming_requests}
Requests arrive with a negative exponentially distributed interarrival time, with rate 1 at the load balancer. Let $I(t)$ indicate that a request arrived at time $t$. Requests have a unique arrival time. Then $I:2^{\mathcal{R}}$ is a infinite set with the arrival times of all incoming requests. For illustration, the following $I$ has been generated using a random number generator:
\vspaceaboveformulas\begin{equation}
\begin{split}
I = \{ 0.87,\:	0.91,\:	1.46,\:	2.03,\:	3.54,\:	4.68,\:	5.42,\:	5.52,\:	5.66,\:	7.26,\:	9.61,\:	10.34,\: 10.93,\:	11.65,\: \cdots \}.
\end{split}
\end{equation}\vspacebelowformulas

\noindent The incoming requests are inspected by the load balancer, which distributes the requests over selected servers for processing, based on a policy. Let $\mathcal{P}(t): \mathcal{R}^+ \rightarrow \{ S_1, S_2, S_3, S_4 \}$ be a load balancer policy that distributes the request that arrived at time $t$ to either server $S_1$, $S_2$, $S_3$ or $S_4$. Let $S_m = \{ t \in I \vert \mathcal{P}(t)=S_m \}$ be the incoming requests of server $m$. Hence, $\{S_1, S_2, S_3, S_4\}$ is a partition of $I$, For illustration, assume policy $\mathcal{P}$ distributes the above incoming requests $I$ over the four servers, as follows.
\vspaceaboveformulas\begin{equation}
\begin{split}
S_1 = \{ 0.91	,\:		3.54	,\:	5.42	,\:				10.34	,\:			
 \cdots\}, & \quad\quad\quad
S_2 = \{ 									9.61,\:						13.04,\:	13.52	,\:
 \cdots\}, \\
S_3 = \{	2.03,\:		4.68	,\:		5.66,\:	7.26,\:			
 \cdots	\}, & \quad\quad\quad
S_4 = \{ 0.87	,\:	1.46	,\:				5.52	,\:						12.01,\:
 \cdots \}.\\
\end{split}
\end{equation}\vspacebelowformulas

\vspaceabovesection
\subsubsection{Power state transitions and energy consumption}
\label{sec:formal_model_performance_energy_power_states}
Servers are in exactly one power state at a time, viz., $on$, suspend ($sd$), sleep ($sl$), or wakeup ($wu$):
\vspaceaboveformulas\begin{equation}
\begin{split}
  P^{on}_x(t) \oplus P^{sd}_x(t) \oplus P^{wu}_x(t) \oplus P^{sl}_x(t),
\end{split}
\label{eq:3}
\end{equation}\vspacebelowformulas

\noindent where $P^s_x(t)$ indicates that resource $x$ is in state $s$ at time $t$.\\

\noindent The current power state of a server is implicitly determined by the amount of incoming requests it receives and how quickly it processes them. E.g., when a server receives less incoming requests, it is more likely to go to sleep to save energy.

\begin{figure}
\centering
\includegraphics[scale=0.8, page=4, trim=0mm 90mm 120mm 0mm, clip]{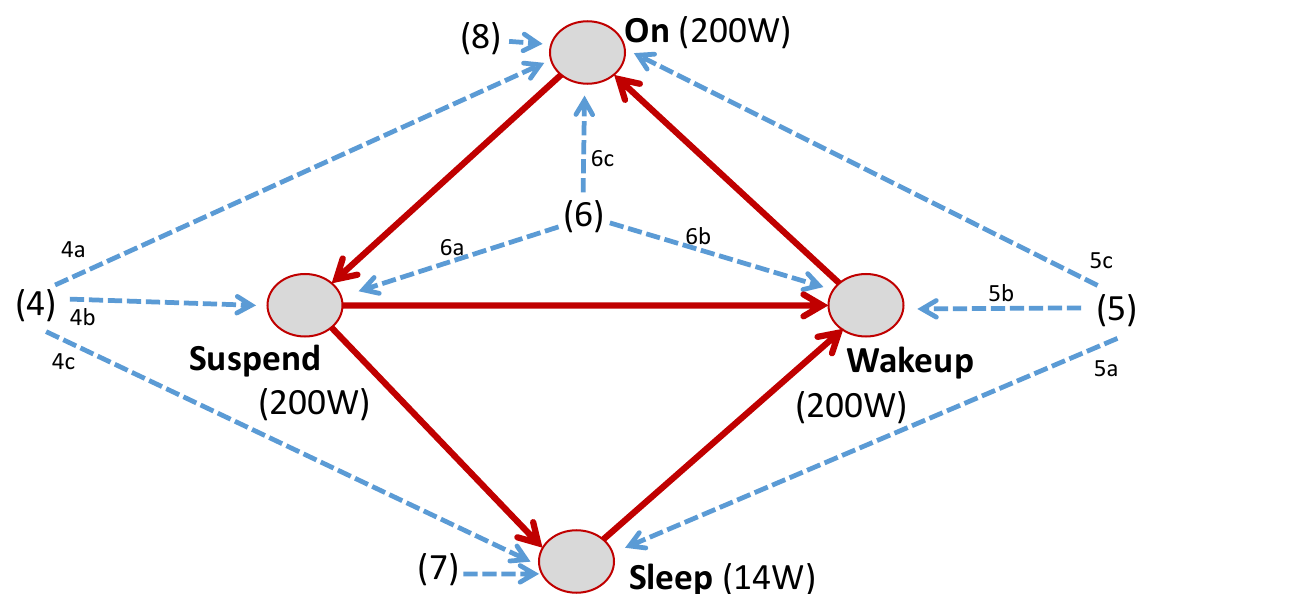}
\caption{Power states of resources and their transitions}
\label{fig:power_states_and_transitions}
\end{figure}

Next, we present the specification of a load balancer with respect to power state behaviour in equations~(\ref{eq:4}-\ref{eq:8}). Figure~\ref{fig:power_states_and_transitions} visualizes the four power states as well as the roles the equations play with them, i.e., equation~(\ref{eq:4}-\ref{eq:6}) are concerned with power three states, and equations~(\ref{eq:7}) and (\ref{eq:8}) with one power state.

When a server finishes processing its last request and when no new requests arrive in the next $TO$ seconds, the server stays on for TO seconds (\ref{eq:4}a), suspends for the next 10 seconds,  (\ref{eq:4}b), and ends in sleep mode (\ref{eq:4}c), as follows.
\vspaceaboveformulas\begin{equation}
\begin{split}
( QS_x(t)=1 \: \land \: QS_x(t+\epsilon)=0 \:\:\:\: \land \:\:\:\: S_x \cap  [t:t+TO] = \emptyset )
  \rightarrow \\
(\underbrace{t'\in[t:t+TO] \rightarrow P^{on}_x(t')}_{\ref{eq:4}b} \:\:\:\: \land \:\:\:\: 
\underbrace{t'\in[t+TO:t+10+TO] \rightarrow P_x^{sd}(t')}_{\ref{eq:4}b}),
\:\:\:\: \land \:\:\:\: 
\underbrace{P_x^{sl}(10+TO+\epsilon)}_{\ref{eq:4}c}),
\end{split}
\label{eq:4}
\end{equation}
\noindent where $TO$ is the time of inactivity needed before a server goes to sleep, $QS_n(t)$ is the queue size plus the request receiving service (either 0 or 1) of server $n$ at time $t$, and $P_x(t)$ the power state (either on, sleep, suspend, wakeup) of server $x$ at time $t$.\\

\noindent When a server is in sleep mode (\ref{eq:5}a) and a request arrives, it wakes up for 10 seconds (\ref{eq:5}b) and then turns on (\ref{eq:5}c), as follows.
\vspaceaboveformulas\begin{equation}
\begin{split}
(QS_x(t)=0 \: \land \: QS_x(t+\epsilon)=1 \:\:\:\: \land \:\:\:\: P^{sl}_x(t) )
 \rightarrow \\
(\underbrace{t'\in[t:t+10] P^{wu}_x(t')}_{\ref{eq:5}b} \:\:\:\: \land \:\:\:\:\
\underbrace{t'\in[t+10:t+10+TO] \rightarrow P^{on}_x(t')}_{\ref{eq:5}c}).
\end{split}
\label{eq:5}
\end{equation}
\noindent When a server is suspending (\ref{eq:6}a) and a request arrives, it will finish suspending and then directly wake up again (\ref{eq:6}b), as follows.
\begin{equation}
\begin{split}
(P^{on}_x(t) \: \land \: P^{sd}_x(t+\epsilon) \:\:\:\: \land \:\:\:\: \exists_{t'\in[t:t+10]} QS_x(t'))
 \rightarrow \\
(\underbrace{t'\in[t:t+10] \rightarrow P^{sd}_x(t')}_{\ref{eq:6}a} \:\:\:\: \land \:\:\:\: \underbrace{t'\in[t+10:t+20] \rightarrow P^{wu}_x(t')}_{\ref{eq:6}b})
\end{split}
\label{eq:6}
\end{equation}

\noindent When a server is in sleep mode (\ref{eq:7}), it remains there as long as no new requests arrive, as follows.
\vspaceaboveformulas\begin{equation}
\begin{split}
P^{sl}_x(t) \:\:\:\: \land \:\:\:\: QS_x(t+t')=0 \rightarrow P^{sl}_x(t+t')
\end{split}
\label{eq:7}
\end{equation}\vspacebelowformulas

\noindent When a server is turned on (\ref{eq:8}), it remains turned on when incoming requests arrive frequently: 
\vspaceaboveformulas\begin{equation}
\begin{split}
P^{on}_x(t) \: \land \: QS_x(t+t')>0 \:\:\:\: \land \:\:\:\: t'<TO \rightarrow P^{on}_x(t+t'),
\end{split}
\label{eq:8}
\end{equation}\vspacebelowformulas

\noindent where $TO$ is the time of inactivity needed before a server goes to sleep. $TO$ is design dependent.

\vspaceabovesection
\subsubsection{Effectiveness of a load balancer}
\label{sec:formal_model_performance_energy_power_latencies}
Section~\ref{sec:effectiveness_load_balancing_policy} addressed performance and energy consumption as the properties to evaluate a load balancer on. The following two equations define, respectively, the performance and energy consumption formally.
Performance is defined as the average latency ($AL$) of all requests conceivable, as follows.
\vspaceaboveformulas\begin{equation}
\begin{split}
AL =  \lim_{n \rightarrow \infty} \:\:\:\:  \frac{1}{n}\sum_{i=1}^{n}L_i,
\end{split}
\end{equation}\vspacebelowformulas

\noindent where $L_i$ is the latency of the $i^{th}$ request.

\noindent Average power consumed ($AP$) to keep all four servers running is defined, as follows.
\begin{equation}
\begin{split}
AP = \lim_{\textit{t} \rightarrow \infty} \sum_{m=1}^4 \frac{1}{t} \int_0^t \frac{200 \cdot (Pi^{on}_m(s) +  Pi^{wu}_m(s) +  Pi^{sd}_m(s)) + 14 \cdot Pi^{sl}_m(s)}{4} ds,
\end{split}
\end{equation}

\noindent where $Pi^x_m(t)$ = 1 when $P^{x}_m(t)$, and ${Pi}^x_m(t)$ = 0, otherwise.

\vspaceabovesection
\subsection{A policy specification for load balancers}
\label{sec:formal_model_policy}
A load balancer policy prescribes how a load balancer behaves with respect to distributing incoming requests to servers. Section~\ref{ref:ordering_of_services} defines a policy using a mechanism that orders the servers; Section~\ref{sec:typical_policies} presents some example policies. Finally, Section~\ref{sec:resolve_nondet} provides ways to resolve non-determinism when a policy is ambiguous.

\vspaceabovesection
\subsubsection{A load balancer policy grammar}
\label{ref:ordering_of_services}
Each time an incoming requests arrives, a load balancer has to select one of the servers to delegate this request to. We use the following algorithm to make this decision: 
\begin{itemize}
\item Relevant system state variables are retrieved, e.g., the queue sizes of the servers.
\item The preference of each server is determined via an arithmetic expression that includes system state variables.
\item The incoming request is delegated to the most desirable server, viz., the server with the highest outcome for the arithmetic expression.
\end{itemize}
\noindent Table~\ref{tab:policy_grammar} shows the grammar of a policy expression $\mathcal{P}$.
Policy expressions can be something from the categories $state$, $power$, $time$ or $math$, as follows.

\begin{table}[t]
\centering
\begin{tabular}{|l|}
\hline
$\mathcal{P} = ID \xx numServers \xx queueSize  \xx state \xx power \xx time \xx math $ \\
$state = stateOn \xx stateSleep \xx stateSuspend \xx stateWakeup$\\
$power = powerOn \xx powerSleep \xx powerSuspend \xx powerWakeup$\\
$time = timeWakeup \xx timeSuspend \xx timeOutTime$\\
$math = \mathcal{P}*\mathcal{P} \xx \mathcal{P}+\mathcal{P} \xx \mathcal{P}-\mathcal{P} \xx \mathcal{P}/\mathcal{P} \xx \mathcal{P}\mod \mathcal{P}  \xx INT \xx random \xx dspace(STRING)$\\
 \hline
\end{tabular}
\caption{The grammar of load balancer policy expression $\mathcal{P}$}
\label{tab:policy_grammar}
\end{table}

\begin{itemize}
\item $State$ provides indicator functions to check whether a server is in one of the four states, or not, e.g., when a server is in the $on$ state, $stateOn$ yields 1 and the others 0.
\item $Power$ is used to retrieve the power consumptions of each individual state (see assumption $\mathcal{A}7$). 
\item $Time$ includes $timeWakeup$ and $timeSuspend$, the time it takes for the server to go back and forth between states $on$ and $sleep$, as well as $timeOutTime$, the time of inactivity the server undergoes before going to sleep (see assumption $\mathcal{A}6$).
\item $Math$ provides five recursive functions that combine policies via arithmetic operations to create arbitrarily complex polices. Furthermore, a constant integer number, a random number $r\in[0:1]$, and a design dependent number could be used.
\end{itemize}

\noindent Furthermore, policy expressions can be an $ID$, a unique number for each server, $numServers$, the total number of servers, and $queueSize$, the number of requests in the queue of each server.

\vspaceabovesection
\newpage
\subsubsection{Example policies for load balancers}
\label{sec:typical_policies}
Let $\mathcal{P}_{q}$ be a generic policy, where $q \geq 0$ is a variable, to illustrate the functioning of policies in practice:
\vspaceaboveformulas\begin{equation}
\mathcal{P}_{q} = -queueSize - q \cdot (1-stateOn),
\label{eq:policy_general}
\end{equation}\vspacebelowformulas

\noindent where $q$ is the server queue size at which an additional server is switched on; $q$ is design dependent.

Then, policy $\mathcal{P}_{0}$ assigns incoming requests to the server that currently has the shortest queue size:
\vspaceaboveformulas\begin{equation}
\mathcal{P}_{0} = -queueSize
\end{equation}\vspacebelowformulas

Table~\ref{tab:outcomes_policy_p0} shows an example evaluation of $\mathcal{P}_{0}$, where $lb_{select}$ is the choice of the load balancer for a certain server, \#n incoming request $n$, and the numbers in the table the policy evaluations of servers 1-4 for incoming request \#n. For the sake of simplicity, we assume that no incoming request finished processing (yet). For request \#1-\#4, the load balancer arbitrary selects servers, because multiple servers have the highest value for the policy evaluation. In step \#5-\#8 this pattern repeats. Hence, requests are equally distributed over the servers.

Below, $\mathcal{P}_{5}$ is also policy that primarily assigns new incoming requests to the server with the shortest queue size. However, $\mathcal{P}_{5}$ also considers the power state of the servers to save energy. Concretely, it will only switch on a new server if all currently switched-on servers have a queue size of at least 5. Note that this policy might perform less well than $\mathcal{P}_{0}$, however, at the benefit of reduced energy consumption.
\vspaceaboveformulas\begin{equation}
\mathcal{P}_{5} = queueSize - 5 \cdot (1-stateOn)
\end{equation}\vspacebelowformulas

\vspace{1cm}
Table~\ref{tab:outcomes_policy_p5} show an example evaluation of $\mathcal{P}_{5}$. When request \#1 arrives, the policy evaluation of all servers yields -5 because no servers are turned on. When the load balancer arbitrarily delegates this request to server 1, server 1 switches on and its policy evaluates to -1. Consequently, the load balancer selects server 1 for request \#2-5. Then request \#6-10 are delegated to server 2 for similar reasons. Note that after 10 incoming requests only two servers have received requests, which is a good property when energy consumption is of concern.

\begin{table}
\centering
\subfloat[performance-optimizing policy ${P}_{0}$]{ \label{tab:outcomes_policy_p0}
\begin{tabular}{ | c | c | c | c | c | c | c | c |  }
\hline
 & \textbf{\#1} & \textbf{\#2} & \textbf{\#3} & \textbf{\#4} & \textbf{\#5} & \textbf{\#6} & \textbf{\#7} \\
 \hline
0 & 0 &	0 &	0 &	0 & -1  & -1 & -1  \\
1 & 0 &	0 &	-1 &	-1 & -1  & -2  & -2 \\
2 &	0 &	0 &	0 &	-1 & -1  & -1 & -2 \\
3 &	0 &	-1 & -1 &	-1 & -1  & -1 & -1 \\
\hline
$lb_{select}$ &	3    &	1    & 2 & 0 & 1 & 2 & 3	 \\
\hline 
\end{tabular}}\quad
\subfloat[reasonably energy-efficient policy ${P}_{5}$]{ \label{tab:outcomes_policy_p5}
\begin{tabular}{ | c | c | c | c | c | c | c | c | c | c }
\hline
 & \textbf{\#1} & \textbf{\#2} & \textbf{\#3} & \textbf{\#4} & \textbf{\#5} & \textbf{\#6} & \textbf{\#7} & \textbf{\#8} \\

\hline
0	 & -5 &	-5 & -5 &	-5 &	-5 & -5 &	-5  & -5\\
1	& -5 &	-1 & -2 &	-3  & -4 & -5 & -5 & -5 \\
2	& -5 &	-5 & -5 &	-5 &	-5 & -5 &	-1 & -2 \\
3	& -5 &	-5 & -5 &	-5 &	-5 & -5 &	-5 & -5 \\
\hline
$lb_{select}$	& 1	 & 1 & 1	 & 1 &1 & 2 & 2 & 2 \\
\hline
\end{tabular}}
\label{tab:outcomes_policy}
\caption{Example evaluations of policies $\mathcal{P}_{0}$ and $\mathcal{P}_{5}$, respectively}
\end{table}

\vspaceabovesection
\subsubsection{Resolving non-determinism}
\label{sec:resolve_nondet}
In this section, we provide a solution that prevents an arbitrary selection of a server by the load balancer when multiple servers have the highest value for their policy expression.

Table~\ref{tab:outcomes_policy_p5} shows an example evaluation for $\mathcal{P}_{5}$. For each incoming request \#1-\#8, the load balancer selects the server with the highest evaluated value, e.g., for request \#2 server 1 gets selected because its policy expression evaluates to -1, which is higher than the -5 of the other three servers. However, there are cases in which the expression of multiple servers has the highest evaluation, e.g., for request \#1 all servers evaluate to 1, which makes selecting server 1 an arbitrary decision. In these cases, the load balancer performs a so-called non-deterministic decision.

Non-determinism can be resolved by adding fractions $f \in [0:1)$ to policy outcomes, as follows. Let $\mathcal{P}$ be a policy that only returns whole numbers $\mathcal{N}$, then policies
\vspaceaboveformulas\begin{align}
\label{eq:nondeterminism}
\mathcal{P}'_{q}=\mathcal{P}_{q} + random &&
\mathcal{P''}_{q}=\mathcal{P}_{q} + \frac{ID}{numServers}
\end{align}\vspacebelowformulas

\noindent yield unique real numbers $\mathcal{R}$ for each server, eliminating non-determinism. The policy outcomes are only unique, if we assume that randomly drawn numbers are unique and if each server has a unique $ID$.

\begin{table}
\centering
\subfloat[random policy $\mathcal{P}' = random$]{\label{tab:nondet_resolve_random}
\begin{tabular}{ | l | l | l | l | l |}
\hline
 & \textbf{\#1} & \textbf{\#2} & \textbf{\#3} & \textbf{\#4} \\
 \hline
0: random & 0.61 &	0.78 &	0.05 &	0.68 \\
1: random & 0.46 &	0.09 &	0.22 &	0.79 \\
2: random &	0.70 &	0.12 &	0.93 &	0.15 \\
3: random &	0.76 &	0.39 &	0.51 &	0.66 \\
\hline
$lb_{select}$ &	3    &	0    &	2    &	2 \\
\hline 
\end{tabular}}\quad
\subfloat[policy fixed order $\mathcal{P}'' = \frac{ID}{numServers}$]{\label{tab:nondet_resolve_fixed}
\begin{tabular}{ | l | l | l | l | l |}
\hline
 & \textbf{\#1} & \textbf{\#2}  & \textbf{\#3} & \textbf{\#4} \\
\hline
0:ID/numservers	 & 0 &	0 & 0 &	0 \\
1:ID/numservers	& 0.25 &	0.25 & 0.25 &	0.25 \\
2:ID/numservers	& 0.5 &	0.5 & 0.5 &	0.5 \\
3:ID/numservers	& 0.75	& 0.75 & 0.75	& 0.75 \\
\hline
$lb_{select}$	& 3	 & 3 & 3	 & 3 \\
\hline 
\end{tabular}}
\caption{Example evaluations of non-determinism resolution mechanisms to two policies}
\label{tab:nondet_resolve}
\end{table}

We illustrate how these two resolution mechanisms work by providing two example evaluations for them, respectively. For the sake of simplicity, we use policies $\mathcal{P}'. = random$ and $\mathcal{P}'' = \frac{ID}{numServers}$.

Table~\ref{tab:nondet_resolve_random} shows how $\mathcal{P}'$ functions. For each incoming request, four random numbers are drawn and the load balancer delegates the request to the server with the highest value, e.g., request \#1 is delegated to server 3 because $max(0.61, 0.46, 0.70, 0.76)=0.76$.

Table~\ref{tab:nondet_resolve_fixed} shows how $\mathcal{P}''$ functions. For each incoming request, the policy evaluates to a unique number per server, which is divided by 4, the number of servers, to return number in range $[0,1)$. The load balancer delegates all requests to server 3 that has the highest ID, namely 3.

\vspaceabovesection
\subsubsection{Design space}\label{sec:designspace}
We define a design space to compare many policies. Each design then represents a load balancer with a different policy. The design space is defined as the Cartesian product over the following three dimensions and their ranges.
\begin{itemize}
\item $q \in \{1,2,3,5,7,10,15,  20,  30,  40,  50,  75,  100\}$ (cf. equation~\ref{eq:policy_general}).
\item timeout time: $TO \in \{ 1,  2,  3,  4,  5,  7.5,  10,  15,  30  \}$ (cf. equation~\ref{eq:8}).
\item non-determinism resolution: $nd \in  \{ random, fixed\_order \}$  (cf. equation~\ref{eq:nondeterminism}).
\end{itemize}
E.g., design $d = (5,10,random)$ is a design in which: (i) a new server is turned on when the queue sizes of all currently running servers is greater than 5, (ii) servers shut down after 10 seconds of inactivity, and (iii) non-determinism is resolved via a random selection.
\newpagebeforechapter

\vspaceabovesection
\section {Two implementations}
\label{sec:two_implementations}
In this section we implement the performance and energy model in two different ways to enable the automatic evaluation of many load balancing policies and validate the results. Section~\ref{sec:iDSL_implementation} provides an iDSL implementation, whereas Section~\ref{sec:anylogic_implementation} provides an AnyLogic implementation.

\vspaceabovesection
\subsection{iDSL implementation}
\label{sec:iDSL_implementation}
iDSL \cite{Bergvaluetools,BergHHHR15,so94137,BergRH15} is a formal language and solution chain to evaluate the performance of service-oriented systems. iDSL has been developed made using Eclipse for DSLs\footnote{http://www.eclipse.org/downloads/packages/eclipse-java-and-dsl-tools/junosr1-rc2} and is, therefore, an Eclipse plug-in with an extensive IDE.
iDSL supports both simulation and model checking, via a transformation to Modest \cite{modest}, as means to evaluate large numbers of complex designs. Finally, iDSL presents its predictions intuitively via visualizations and understandable (aggregated) metrics.

We extend iDSL to support load balancers, as follows. In Section~\ref{sec:idsl_language}, we provide an overview of the iDSL language. In Section~\ref{sec:idsl_implement}, we extend the iDSL language with a load balancer construct and define its semantics via a transformation to Modest. In Section~\ref{sec:idsl_instance}, we define an iDSL instance with a load balancer.

\begin{figure}
    \centering\includegraphics[scale=0.72]{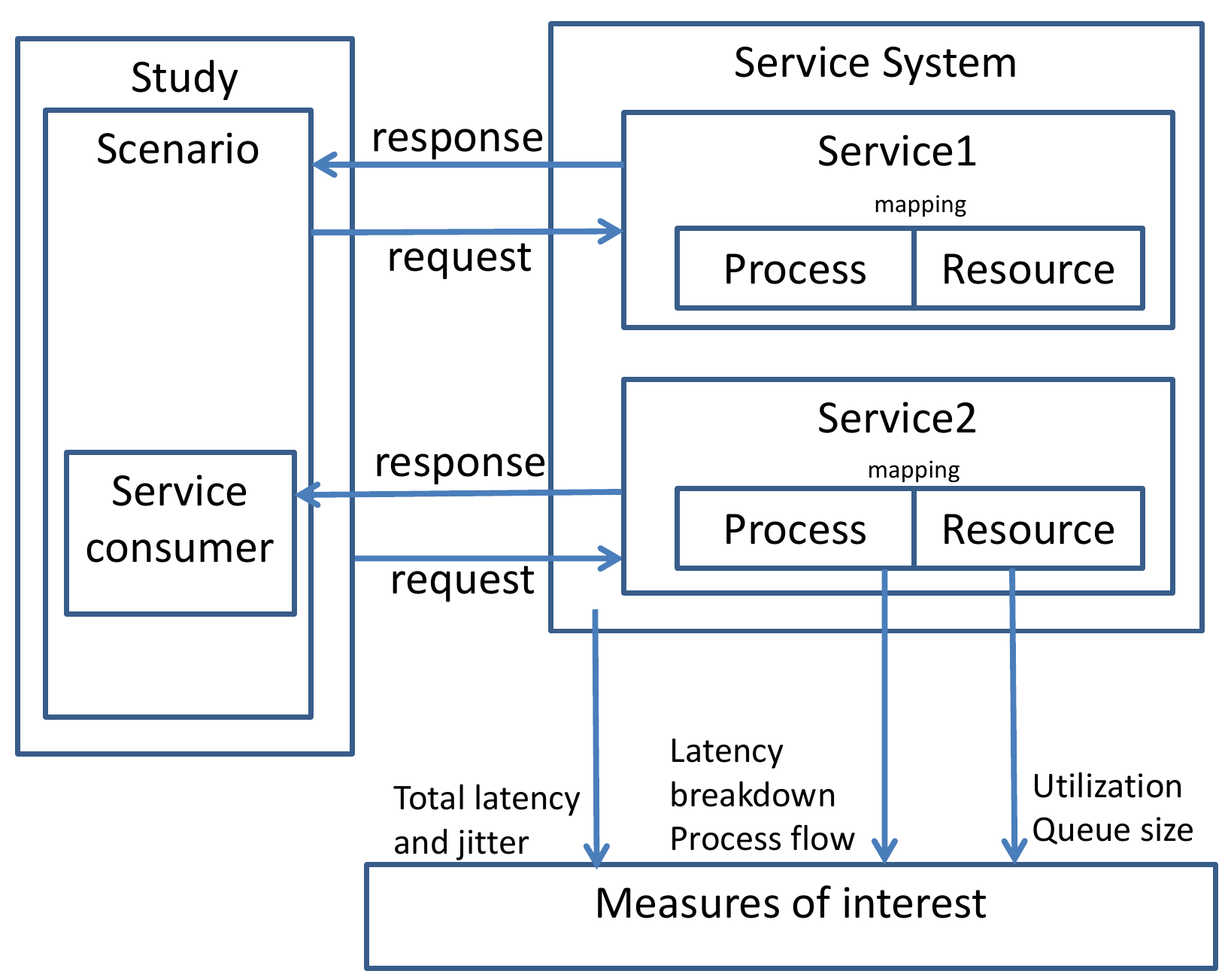}
    \caption{Conceptual model of a service system. Measures of interest are obtained using scenarios.}
    \label{fig:service_system}
\end{figure}

\subsubsection{The iDSL language}
\label{sec:idsl_language}
We describes the conceptual model that forms the basis of iDSL (as depicted in Figure~\ref{fig:service_system}).

A \textbf{service system}, as depicted in the upper right block, provides services to service consumers in its environment, exterior to the service system. A consumer can send a request for a specific service at a given time, after which the system responds after some delay.

A \textbf{service} is implemented using a process, resources and a mapping. \textbf{process} decomposes high-level service requests into atomic tasks, each assigned to resources through the \textbf{mapping} (from which we abstracted in the figure). Hence, the mapping forms the connection between a process and the resources it uses. \textbf{Resources} are capable of performing one atomic task at a time, in a certain amount of time. When multiple services are invoked, their resource needs may overlap, causing concurrency and making performance analysis more challenging.

A \textbf{scenario} consist of a number of invoked service requests over time  to observe the performance behaviour of the service system in specific circumstances. We assume service requests to be functionally independent of each other. That is, service requests do not affect each other's functional outcomes, but may affect each other's performance implicitly.

A \textbf{study} evaluates a selection of systematically chosen scenarios to derive the system's underlying characteristics. Finally, \textbf{measures of interest} define what performance metrics are of interest, given a system in a given scenario. Measures can either be external to the system, e.g., throughput, latency and jitter, or internal, e.g., queue sizes and utilization.

\vspaceabovesection
\subsubsection{Extending iDSL to support load balancers}
\label{sec:idsl_implement}
We extend iDSL to support load balancers. A language construct named $lbalt$ is generated first (see Figure~\ref{fig:ids_process_load_balancer}, Section Process). $lbalt$ contains a \textit{policy} (as defined in Section~\ref{ref:ordering_of_services}), a \textit{configuration} with power consumptions per state and transition times (of Section~\ref{sec:formal_model_performance_energy_power_states}), and multiple processes that are mapped to resources to distribute incoming request to. Also, iDSL has been extended to support energy-efficient resources with four power states.

Under the hood, the load balancer and energy-efficient resources are transformed to multiple Modest \cite{modest} processes, as follows. Figure~\ref{fig:modest_processes} shows initial  process $main$ that initializes a \textit{load balancer} process, a \textit{generator} for incoming requests, and four energy-efficient \textit{resources}, in parallel. To load balancer becomes active when an \textit{incoming request} enter the system, viz., the load balancer evaluates the policy for each resource and adds the incoming request to the buffer of the preferred resource. At the same time, resources wait for a request to  arrive in the buffer, which they then process. Alternatively, the resource times out: it goes to \textit{suspend} mode (in process \textit{resource\_off}) first, then to \textit{sleep} mode, and finally waits for a request to arrive in its buffer. When this happens, it turns \textit{on} again (by calling process \textit{resource}). Note that the power consumption has been modelled using a reward named power.

\vspaceabovesection
\subsubsection{IDSL instance of the load balancer and experiments}
\label{sec:idsl_instance}
\begin{figure}
\centering
\includegraphics[scale=0.55, page=3, trim=0mm 60mm 0mm 0mm, clip]{images/paper_graphs.pdf}
\caption{The Modest processes for the load balancer}
\label{fig:modest_processes}

\centering
\begin{framed}\includegraphics[scale=0.59]{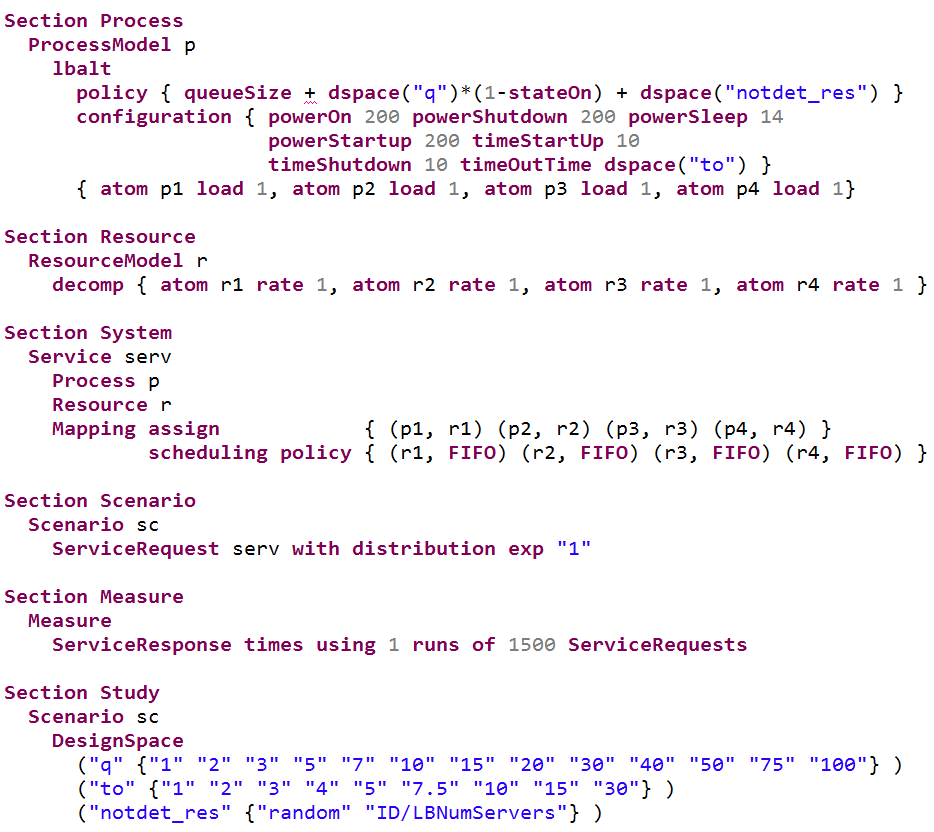}
\end{framed}
\caption{A full iDSL instance for a load balancer with four servers}
\label{fig:ids_process_load_balancer}
\end{figure}

Figure~\ref{fig:ids_process_load_balancer} shows a full iDSL instance of a load balancer with four servers. The iDSL consists of the following six sections. Section \textit{Process} defines a load balancer (as explained in Section~\ref{sec:idsl_implement}) that consists of a policy (based on equation~(\ref{eq:policy_general}), a configuration (based on assumptions $\mathcal{A}6$ and $\mathcal{A}7$), and four processes. In Section \textit{Resource} the four servers are defined. Section \textit{System} then maps the four processes to the four servers, respectively, via a FIFO scheduling policy. In Section \textit{Scenario} the exponential rate of the incoming requests is set to 1. Section \textit{Measure} indicates that simulation runs of 1500 incoming requests each are used. Finally, Section \textit{Study} defines the design space, the ternary Cartesian product of dimensions $q$, $to$, and $nondet\_res$. Note that the variables are used in the the policy and configuration of the load balancer via the $dspace$ construct, which means these vary per design.

\vspaceabovesection
\subsection{AnyLogic implementation}\label{sec:anylogic_implementation}

\begin{figure}[ht!]
\centering
\includegraphics[scale=0.33]{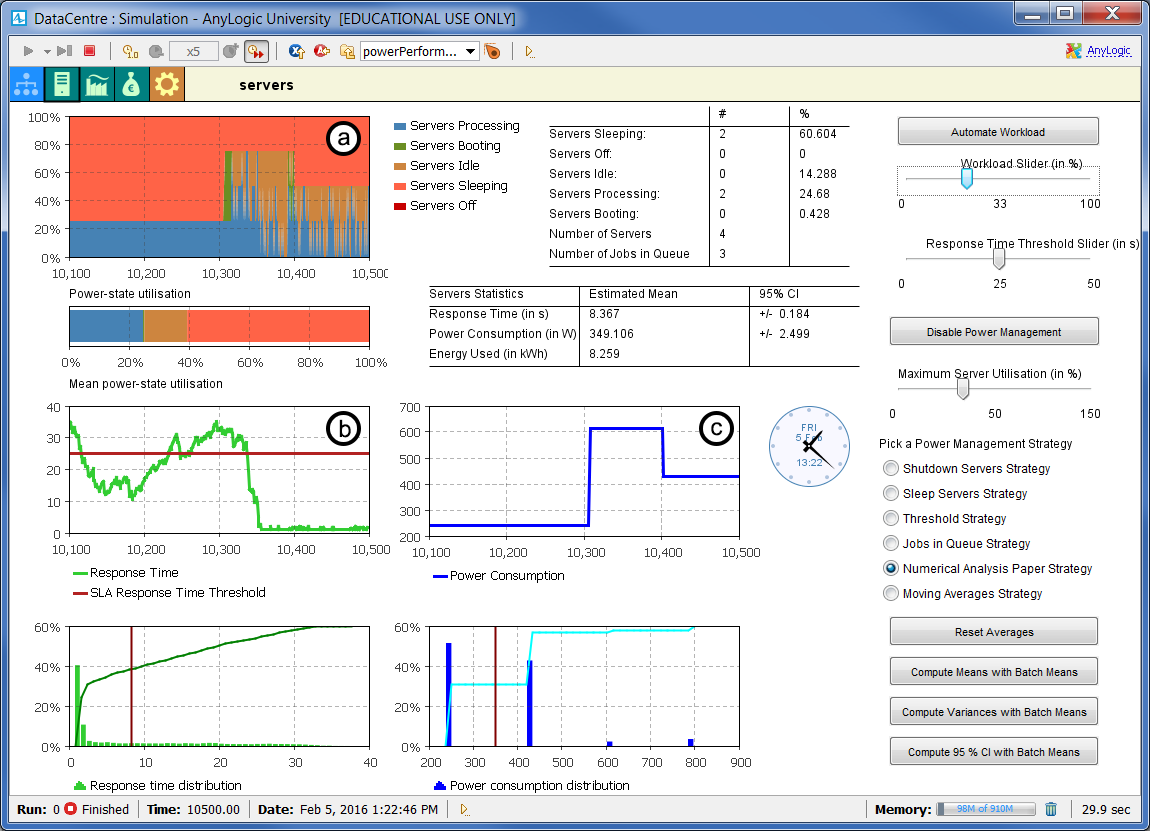}
\caption{The AnyLogic dashboard}
\label{fig:anylogic_dashboard}
\end{figure}

In \cite{Postema}, a simulation framework proposed that allows for the analysis of power and performance trade-offs for data centres that save energy via power management. A combination of high-level models is formed to estimate data centre power consumption and performance. These high-level cooperating simulation models are concerned with (i) IT equipment, (ii) the cascade effect, (iii) the system workload, and (iv) power management. The framework is developed in the AnyLogic \cite{AnyLogic2000} multimethod simulation software, which allows the use of a combination of discrete-event and agent-based models. The framework offers an intuitive dashboard to actively control and obtain insight during each simulation run, as illustrated in Figure~\ref{fig:anylogic_dashboard}. Besides insight into transient behaviour, as can be seen in this figure for the (a) \textit{power-state utilisation}, (b) \textit{response times} and (c) \textit{power consumption}, also averages are computed and depicted in tables to give an indication of the steady-state behaviour.

In Section \ref{sec:config}, the configuration of the data centre in the simulation framework is elaborated. An extension for policies of load balancing and power management for this framework is introduced in Section \ref{sec:policies}. 

\vspaceabovesection
\subsubsection{Configuring the simulation framework}\label{sec:config}

The framework is configured according to the system description (of Section \ref{sec:system_description}). The basic load balancer and its environment are implemented with one agent for the load balancer and one agent for each server. The load balancer distributes the workload by injecting jobs to the servers. These jobs are injected in simple queues inside the server agents.

\vspace{10px}

Table \ref{tab:anylogic_config} shows all the parameters used. Jobs arrive in the load balancer according to a Poisson process with rate ($\lambda$ job/s). The service rate ($\mu$ job/s) of each server is deterministic, i.e., the server finishes jobs in a fixed amount of time. The model is extended to support four power states, cf. (\ref{eq:3}). The awareness of power states in the models allows to compute power consumption ($P$) by rewarding each power state with a power consumption. Note that processing is the only power state in which jobs are served.

\vspacebelowtables\begin{table}[ht!]
\centering
\subfloat[power management parameters]{
\begin{tabular}{| l | l | l | l |}
	\hline
	$P_{pc}$ & 200\,W & $P_{as}$ & 14\,W \\
	\hline
	$P_{sl}$ & 200\,W & $P_{sl}$ & 200\,W \\
	\hline
\end{tabular}}\quad\quad\quad
\subfloat[performance parameters]{
\begin{tabular}{| l | l | l | l |}
	\hline
	$\lambda$ & exp(1.0) & $\mu$ & det(1.0) \\
	\hline
	$\alpha_{sl}$ & det(10.0) & $\alpha_{wk}$ & det(10.0) \\
	\hline
\end{tabular}}
\\
\caption{Data centre configuration parameters}\label{tab:anylogic_config}
\end{table}\vspacebelowtables

\noindent The mean power consumption ($E[P]$) and mean response times ($E[R]$) are computed using the batch means method. The batch means method requires the length of the simulation ($t_{sim}$) to be very long, which is usually around 100,000 virtual seconds, and the system should be stable after some warm-up ($wup$) period, which is usually around 500 virtual seconds.

\vspaceabovesection
\subsubsection{Policy implementation}\label{sec:policies}
Recall that the load balancer injects jobs in the queues of agents of the servers. Therefore, the load balancer needs a policy to determine where each job should go. The policies, introduced in \cite{Postema}, have two simple options: (i) random and (ii) shortest queue next. So, the load balancer required an extension to support policies (cf. Section \ref{sec:policies}).

In order to implement these policies, information is required about the size of the queue of each server and about the current power state. In order to select a server that support these policies an expression should be defined to reward each server. The extension consist of a module class that has access to all the relevant information, such that it can rate the servers based on the three parameters (cf. Section \ref{sec:designspace}). Additionally, a parameter variation experiment of the framework is implemented that allows for parallel computation of the averages of many designs.


\newpagebeforechapter
\vspaceabovesection
\section{Experimental results}\label{sec:expresults}
We show what results of the evaluated designs tell us about policies (in Section~\ref{sec:results_lessons_learned}) and their validity (in Section~\ref{sec:validity}).

\vspaceabovesection
\subsection{Lessons learned}
\label{sec:results_lessons_learned}
\begin{figure}
	\centering
    \subfloat[The value $q$ determines the amount of power traded for performance.]{\includegraphics[scale=0.30]
    {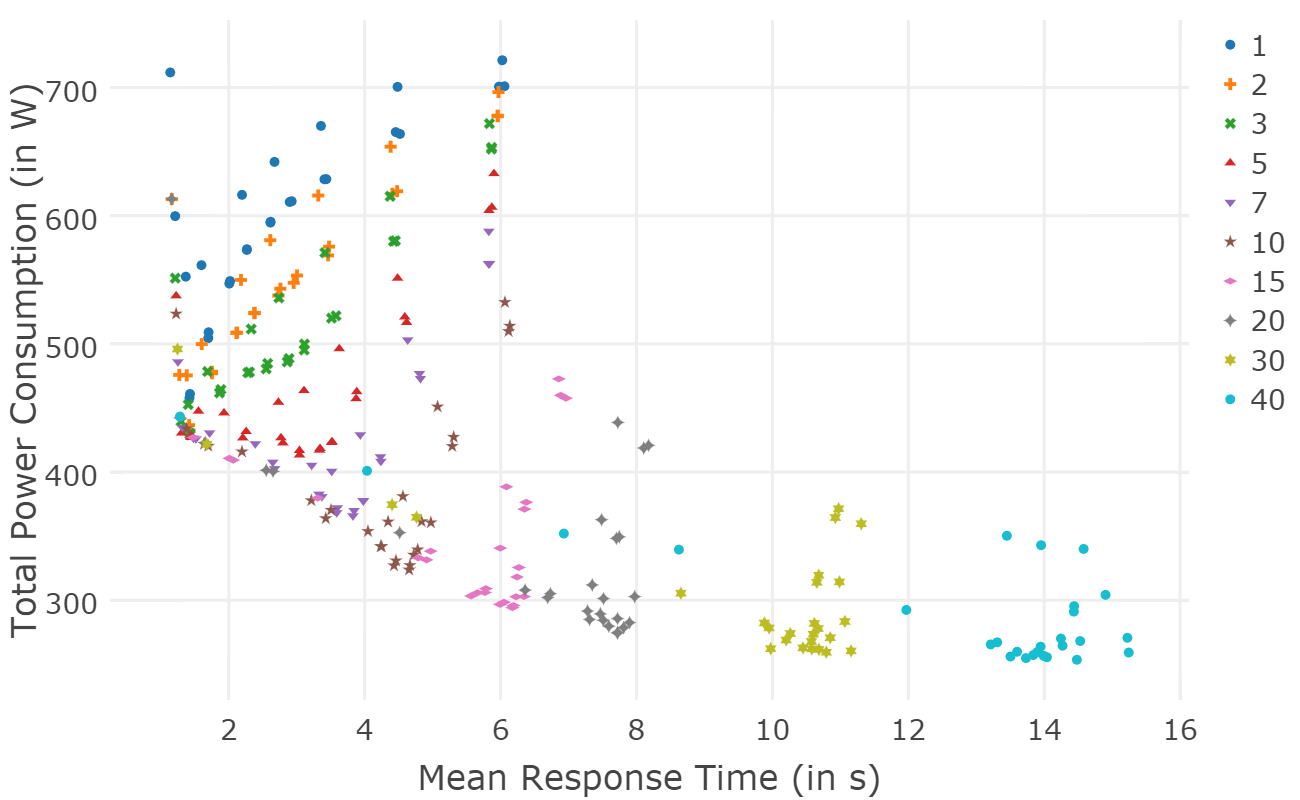}
    \label{fig:qtrading}}\\\vspace{4mm}
  \subfloat[Time-out $to$ determines the position of the frontier.]{\includegraphics[scale=0.30]
  {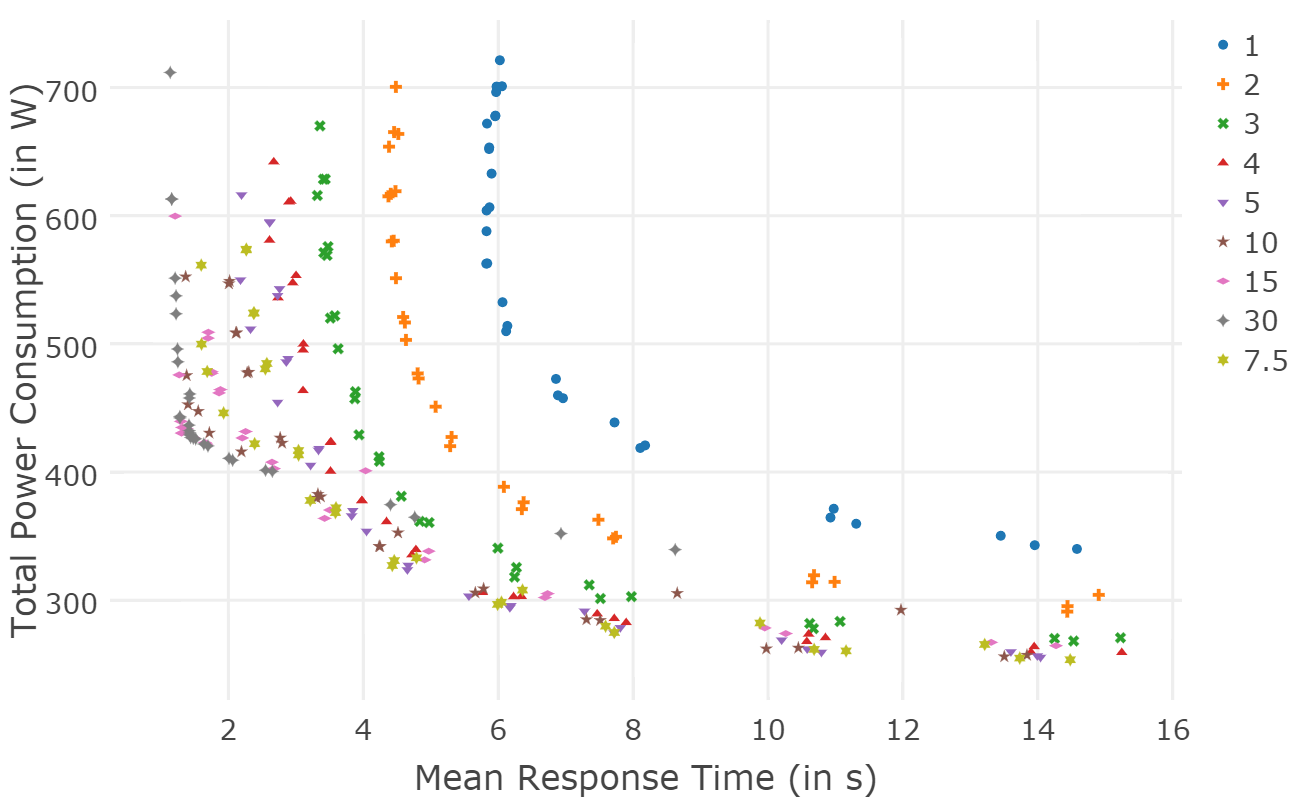}
  \label{fig:tofrontier}}
  \caption{Average latency and power consumption outcomes for many designs}
  \label{fig:qtrading_and_tofrontier}
\end{figure}

Figure~\ref{fig:qtrading_and_tofrontier} shows the Anylogic results for all designs. It shows the effect of adjusting the parameters in the policy for various values of queue threshold $q \in \{1,2,3,5,7,10,15,20,30,40\}$ and the time-out $TO \in \{1,2,3,4,5,7.5,10,15,30\}$. 
In Figure~\ref{fig:qtrading} equal values for $q$ are marked with the same colour. It shows that high values of $q$ lead to low energy consumption and low values of $q$ to high performance. Figure~\ref{fig:tofrontier} has similar colours $TO$. It shows that the efficiency frontier depends on the time-out value $TO$.

\begin{figure}
	\centering
    \includegraphics[scale=1, page=1]{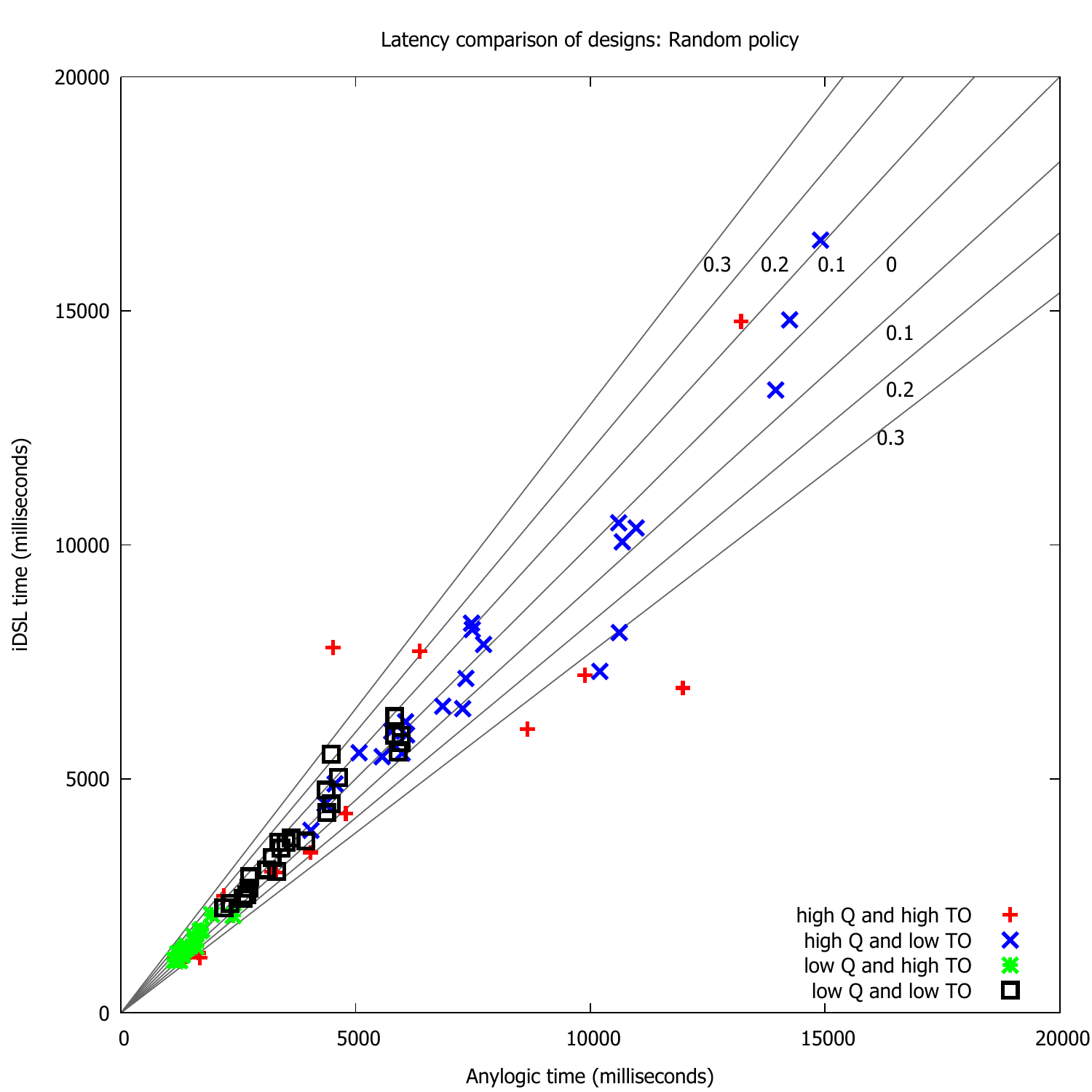}
  \caption{Comparison of the iDSL (on the $y$-axis) on AnyLogic (on the $x$-axis) results for many designs, using random non-determinism elimination}
    \label{fig:idsl_anylogic_policy0}
\end{figure}

\begin{figure}
	\centering
  \includegraphics[scale=1, page=2]{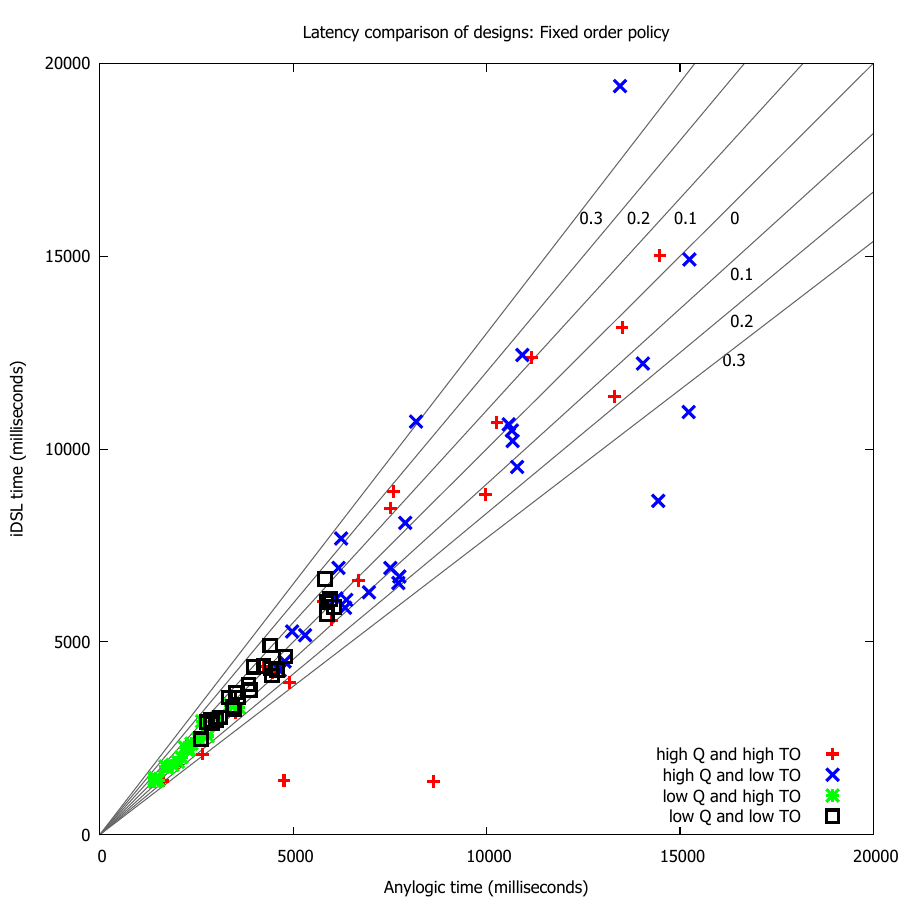}
  \caption{Comparison of the iDSL (on the $y$-axis) on AnyLogic (on the $x$-axis) results for many designs, using fixed order non-determinism elimination}
  \label{fig:idsl_anylogic_policy1}
\end{figure}



\vspaceabovesection
\newpage
\subsection{Validity of the outcomes}
\label{sec:validity}
We assess the validity of the iSDL and AnyLogic approaches by comparing their performance and energy consumption outcomes for many different designs. The following distance measure, which returns the ratio differences, is used to compare outcomes:
\begin{equation}
\delta (v_1,v_2) = \max\left(\frac{v_1}{v_2},\frac{v_2}{v_1}\right)-1
\end{equation}
The measure is partly like a metric, viz., $\delta(v,v)=0$, $\delta(v_1,v_2)=\delta(v_2,v_1)$, and $\delta(a v_1,a v_2)=\delta(v_1,v_2)$. However, the triangular property $\delta(v_1,v_2)+\delta(v_2,v_3)\geq\delta(v_1,v_3)$ does not hold.

Figure~\ref{fig:idsl_anylogic_policy0}~and~\ref{fig:idsl_anylogic_policy1} show the experimental outcomes of iDSL (on the $y$-axis) and AnyLogic (on the $x$-axis) for resolving non-determinism with the random (in Figure~\ref{fig:idsl_anylogic_policy0}) or the fixed order (in Figure~\ref{fig:idsl_anylogic_policy1}) way, respectively. Note that the distance $\delta$ is visualised around the diagonal for values 0, 0.1, 0.2 and 0.3. Generally, the results of both implementations match, because most designs are located near the diagonal.



\newpagebeforechapter
\vspaceabovesection
\section{Conclusions}
\label{sec:conclusions_and_future_work}
In this paper we have constructed a model to evaluate the performance and energy consumption of load balancers. In this model, we define a powerful policy language that decide to which server jobs are assigned by  observing the system variables, e.g., queue sizes of servers.

To evaluate the performance and energy trade-off of many policies, we have implemented two load balancers with exactly the same specifications in iDSL \cite{Bergvaluetools,BergHHHR15,so94137,BergRH15} and in AnyLogic \cite{AnyLogic2000}. Alternatives for iDSL, which offers a high-level language, are PRISM \cite{prism}, Modest \cite{modest} and UPPAAL \cite{uppaal}. Cloudsim \cite{calheiros2011cloudsim} is  an alternative for AnyLogic.

Evaluation of many policies shows that parameter $q$, the queue threshold for switching servers on, is useful to resolve the performance and power consumption trade-off, viz., low $q$ values leads to good performance, while higher $q$ values reduce energy consumption. Parameter $TO$, the idling time of servers before sleeping, determines the position of the so-called Pareto optimal frontier. A higher $TO$ value improves both performance and power consumption.

For validation, the evaluated performance and energy consumptions results of both implementations have been compared. For half of all the designs, both the average latencies and power consumption of iDSL and AnyLogic differed less than 6\%. For 80\% of the designs, this is 13\% and 11\%, respectively.

\textit{Related work} The work of \cite{Van2010} simulate models that consider virtual machines and in particular the power-performance trade-off. Similarly, \cite{Petrucci2011} considers virtual machines and a power-performance trade-off with testbed to apply their models for monitoring and control. In \cite{Gandhi}, power management is discussed with a strong focus on server allocation. Furthermore, \cite{Pinheiro} performed on cluster-based systems with a load balancer taking power and performance into account. Finally, \cite{calheiros2011cloudsim} offers power and performance analysis for data centres.

Our work distinguish itself in the following three ways:
First, we have constructed a load balancer policy with a powerful yet concise language which is used to access system variables, such as queue sizes and power states of servers. 
Second, we have implemented this policy in two different development environments, iDSL and AnyLogic. Both implementations were validated by comparison of evaluated results. 
Third, evaluation of many designs provides insight in the meaning of the policies while only using  two parameters: The queue size threshold the affecting the performance power trade-off, the server idle time affecting the level of Pareto optimality.


\vspaceabovesection
\newpagebeforechapter

\bibliographystyle{eptcsini}
\bibliography{library}

\end{document}